\documentclass[12pt]{iopart}

\usepackage{graphicx}
\usepackage{psfrag}
\newcommand{\veps}{{E}}
\newcommand{\Ec}{{\cal E}_{\rm C}}

\begin{document}

\title[Casimir interaction
in the Fermi sea]{Casimir interaction between normal
             or superfluid grains in the Fermi sea}

\author{A Wirzba$^1$, A Bulgac$^2$, P Magierski$^{2,3}$}
\address{$^1$ Institut f\"ur Kernphysik (Th),
              Forschungszentrum J{\"u}lich,
              D-52425 J{\"u}lich, Germany}
\address{$^2$ Department of Physics,
              University of Washington,
              Seattle, WA 98195--1560, USA}
\address{$^3$ Faculty of Physics,
              Warsaw University of Technology,
              %ul. Koszykowa 75, 
              00-662 Warsaw, Poland}
\ead{a.wirzba@fz-juelich.de}

\begin{abstract}
We report on a new force that acts on cavities (literally empty regions of
space) when they are immersed in a background of non-interacting fermionic
matter fields.  The interaction follows from the obstructions to the
(quantum mechanical) motions of the fermions caused by the presence of bubbles
or other (heavy) particles in the Fermi sea, as, for example, nuclei
in the neutron sea in the inner crust of a neutron star or superfluid
grains in a normal Fermi liquid.
The effect resembles the traditional Casimir interaction between metallic
mirrors in the vacuum. However,
the fluctuating electromagnetic fields 
are replaced by fermionic matter fields. We show that
the fermionic Casimir problem for a system of spherical cavities can be
solved exactly, since the
calculation can be mapped onto a quantum mechanical
billiard problem of a point-particle scattered
off a finite number of non-overlapping spheres or disks. Finally we generalize
the map method
to other Casimir systems, especially to the case of
a fluctuating scalar field between two spheres or a sphere and
a plate under Dirichlet boundary conditions.
\end{abstract}

%Uncomment for PACS numbers title message

% Keywords required only for MST, PB, PMB, PM, JOA, JOB? 
%\vspace{2pc}
%\noindent{\it Keywords}: Article preparation, IOP journals
% Uncomment for Submitted to journal title message
%\submitto{\JPA}
% Comment out if separate title page not required
%\maketitle

\section{Introduction}

\subsection{The original Casimir effect}
In 1948 the Dutch physicist H.B.G.\,Casimir predicted the remarkable effect
\cite{casimir48} that
two parallel, very closely spaced, uncharged metallic
plates 
attract each other  in vacuum. 
The origin of this force can be traced back to the altered (mode 
sums over the)
zero-point fluctuations of
the electromagnetic field which are induced by the presence of 
the two plates which are added 
relative to the free case, or rather, which are brought from 
very far separation into a close distance.
The distinctive property of the Casimir effect is that its strength and, 
perhaps, its
sign are geometry-dependent (for a review see~\cite{bordag}).
Our aim is to generalize  two features of
the Casimir effect, the mode sum structure and the geometry-dependence, to the
{\em fermionic Casimir effect}, where the fluctuating photons are replaced by 
(non-relativistic) Fermi fields.

\subsection{Utilizing the geometry dependence of the Casimir energy}
Let us invert the logic  and define the Casimir energy as the energy
resulting from the {\em geometry-dependent} part of the density
of states (d.o.s.) -- a concept that is closely related to the shell
correction energy in nuclear physics:
\begin{equation}
\rho(\veps)
\equiv \sum_{\veps_k} \delta(\veps-\veps_k) = \rho_0(\veps)
+ \rho_{\rm bulk}(\veps) +\delta\rho_{{\rm C}}(\veps,
\mbox{\footnotesize\em geom.-dep.})\,.
\end{equation}
Here $\{E_k\}$ are the eigenenergies of the modes,
$\rho_0$ is the d.o.s.\ of the homogeneous background and
$\rho_{\rm bulk}$ is the
bulk d.o.s.\ that sums up  the excluded volume effects, surface
contributions and Friedel oscillations caused by
each of the obstacles separately. The remaining part
 $\delta\rho_{\rm C}$ is of central interest to us. It is 
the only term which ``knows'' about
the relative geometry-dependence of the obstacles.
Now the Casimir energy can be extracted from the {\em
  geometry-dependent}
part of the density of states as a simple integral
\begin{equation}
 \Ec  \equiv \int d E \, E\, \delta\rho_{\rm C}
(E,\mbox{\footnotesize\em geom.-dep.})%\\ &=&\
=- \int d E\, {\cal N}_{\rm C}(E,\mbox{\footnotesize\em geom.-dep.})\,,
\end{equation}
where ${\cal N}_{\rm C}(E)$ is the geometry-dependent part of the 
integrated density of states 
or number of states,
$
{\cal N}(\veps) \equiv \sum_{\veps_k}
\Theta(\veps-\veps_k)
=\int_{0}^\veps d \veps'\, \rho(\veps')
$.

\subsection{Generalization of the
Casimir energy concept to  matter fields}
Let us assume that space is not ``filled'' with fluctuating electromagnetic
modes, but with a gas of   {\em non-interacting} (non-relativistic) fermions.
Under this scenario we have the following similarities with the ordinary
Casimir effect:
In both cases, there exist mode sums,
$\sum_k \hbar \omega_k$, with
{\em constant}  degeneracy factors.
The constant weight of the fluctuating electromagnetic
modes  can be traced back to
the two helicity states of the photon.  The constancy in the degeneracy of
the fermionic matter modes, on the other hand,
follows from Pauli's exclusion principle, where the pertinent weight
can be formulated  in
terms of spin and isospin factors.
However, the Casimir mode summation over matter fields differs from the case of
fluctuating
fields by the presence of a new independent scale in addition to the geometric
scales (e.g.\ the
separation $L$ and area
$A$ of the plates), namely by the presence of the
Fermi energy (i.e., the chemical potential $\mu$ at zero temperature).

In the following we will consider the case of matter fields
(non-relativistic fermions) located in the space between voids or cavities,
such that the matter fields  will build up a quantum pressure on the voids.
Even if we assume that the matter fields are non-interacting and the voids
are non-overlapping, an
{\em effective} interaction between these 
empty regions of space will still arise
in the background of the non-interacting 
fermionic matter fields, since the cavities
-- depending on their geometric arrangement -- can
shield the free movement of the matter modes such that a net change in the
mode sum over the Fermi states in the Fermi sea results.

Applications of this scenario exist e.g.\ for the inner crust of neutron star.
With increasing distance from the star's surface (inside the star), 
the nuclei start to loose neutrons due to the growing pressure and density.
Bulgac and Magierski argued that the shell correction
energies (in other words the fermionic Casimir energies) 
of the resulting low-density neutron  matter 
in the presence of bubbles (i.e.\ nuclei) are
of the same order of magnitude~\cite{abpm} as the differences  
between the 
usually assumed liquid-drop-model sequence of phases~\cite{folklore}. 
Thus disordered lattices of bubbles are in competition with the standard
sequence of (i) nuclear drops (nuclei), (ii) nuclear rods, (iii) nuclear plates
(all three surrounded by a low-density background of neutrons), 
(iv) tubes, (v) bubbles (both surrounded by the high-density background of 
nuclear matter),
until finally  a  uniform nuclear matter phase is reached.

An analogous investigation of bubbles inside a Fermi gas background is also
of relevance for the inner {\em core}
of neutron stars. Under the assumption
that quark matter does exist  and quark droplets can form there,  
a  similar pattern  is predicted, with 
quark-droplets (bags) inside hadronic matter taking over the role of the
 embedded nuclei.

Also in the laboratory, the study of the interaction of
cavities inside a uniform fermionic background could be 
of importance~\cite{abaw01}.
Examples are C$_{60}$ buckyballs immersed in liquid mercury. The liquid
metal itself serves only as a non-rigid neutral
background which  provides
the Fermi  gas environment via its conductance electrons,
in which the buckyballs ``drill'' the voids.
Another example would be buckyballs in liquid
$^3$He as Fermi gas. Finally, in the future, boson-condensate droplets
immersed
in dilute atomic Fermi condensates could serve as  systems with which
the effective interactions of cavities inside a Fermi gas could be studied in
the lab.

\section{Casimir calculation mapped onto a scattering problem}

Note that the Casimir calculation for fermionic
(non-relativistic) matter
fields simplifies enormously, because the presence of a second scale,
the chemical potential ${ \mu}$=$\hbar^2 k_F^2/2m$
(or the Fermi momentum $k_F$), provides for a
natural $UV$-cutoff,  $\Lambda_{UV} \equiv \mu$ (here $m$ is the  mass of
the fermion).
Thus the Casimir energy for fermions between two impenetrable (parallel)
planes at a distance $L$ is simply given by the 
chemical potential times a finite
function of the dimensionless argument $k_F L$,
$
\Ec = \mu\, F(k_F L)$.

For more complicated geometries, the computations become more and more involved
as it is the case for the ordinary electromagnetic Casimir effect.
However, Casimir calculations of a finite number of
immersed non-overlapping {\em spherical} voids or rods, i.e.\ spheres and
cylinders in 3
dimensions or disks in 2 dimensions, are still doable. In fact, these
calculations simplify because
of Krein's trace formula~\cite{krein,uhlenbeck}
\begin{equation}
\delta\bar\rho(\veps) = \bar\rho(\veps)-\bar\rho_0(\veps)
= \frac{1}{2\pi i} \;\frac{d\, }{d \veps}{\rm tr} \ln S_n (\veps )\,,
\end{equation}
which links the variation in the level density
$\delta\bar\rho(\veps)$ (the difference of the total density of states and the
background one)
to the energy-variation of the phase shift
${\textstyle\frac{1}{2i}}{\ln\det} S_n(E)$ of the
$n$-sphere/disk scattering matrix $S_n(E)$.
Note that the level densities on the left hand side are  averaged
over an energy-interval larger than the mean-level spacing in the volume
$V$ of the entire system in order to match the {\em continuous} expression
on the right hand side.
In this way
the Casimir calculation is mapped to the  quantum mechanical analog of a
classical ``billiard''
problem: the 
hyperbolic or even chaotic scattering of a point-particle off
an assemble of $n$ non-overlapping spheres
(or disks)~\cite{eckhardt,gaspard,cvitanovic,wreport,wh98,hwg97}.

Moreover, it is now possible to extract the  geometry-dependent Casimir 
fluctuations 
from the {\em multiple}-scattering part
of the scattering
matrix. In fact, as shown in \cite{wreport,wh98,hwg97}, 
the determinant of the $n$-sphere/disk S-matrix can be
separated into a product of the determinants of the 1-sphere/disk
S-matrices $S_1(E,a_i)$, where $a_i$ is the radius of the single scatterer $i$,
and the ratio of the determinant of the inverse multi-scattering matrix $M(k)$
and its
complex conjugate:
\begin{equation}
\det{S_n(E)} =
\left\{\prod_{i=1}^n \det{S_1(E,a_i)}\right\}
\frac{\det \left(M(k^*)\right)^\dagger}{\det{ M}(k)}
\, .
\end{equation}
For the fermionic Casimir case, 
the energy $E$ and the wave number $k$ are related as $E=\hbar^2 k^2/2m$.
When inserted into Krein's formula,
the product over the single-scatterer determinants generates
just the bulk (or Weyl term) contribution  to the density of states
\begin{equation}
  \bar\rho_{\rm bulk}(E,\{a_i\})
\equiv \sum_{i=1}^n \bar\rho_{\rm Weyl}(E,a_i)
= \frac{1}{2\pi i}\,\frac{d}{d E}\,
\sum_{i=1}^n \ln \det S_1(E,a_i)\,,
\end{equation}
which takes care of the excluded volume terms and the surface terms (including
Friedel oscillations).
The geometry-dependent part of the d.o.s.\ is therefore given by a modified
Krein equation~\cite{abaw01} which is formulated in terms of the (inverse)
multi-scattering matrix instead of the full S-matrix
\begin{eqnarray}
\delta\bar\rho_{\rm C}
(\veps,\{a_i\},\{{ {\vec r}_{ij}}\}) &=& \bar\rho(E)-\bar\rho_0(E)
-  \sum_{i=1}^n \bar\rho_{\rm Weyl}(E,a_i)\nonumber\\ &=&
-{\frac{1}{\pi }}
{\rm Im}\,
\left\{
 {\frac{d}{d\veps}}
 \ln \det { M}\left[ k(\veps),\{a_i\},\{{ {\vec r}_{ij}}\}\right] \right\}\,,
 \label{modKrein}
\end{eqnarray}
where ${\vec r}_{ij}$ are the relative separation vectors between the
centers of the
spheres (or disks).
The pertinent Casimir energy can then be read off from the finite integral
\begin{equation}
  \Ec = \int_0^\mu d E\, \left( E- \mu\right) \delta\bar\rho_{ C}
= - \int_0^\mu d E\, {\overline{\cal N}}_{\rm C}\,.
\end{equation}
%Note that the chemical potential $\mu$ serves here as  UV cutoff. Thus the
%integral is finite.

\section{The calculation}
Equation (\ref{modKrein}) does allow to simplify the problem, since
there exists a close-form expression for the inverse multi-scattering matrix
for $n$ spheres (of radii $a_j$ and mutual separation 
$r_{jj'}$,
labelled by the  indices $j,j'=1,2,\cdots,n$)
in terms of spherical
Bessel and Hankel functions, spherical harmonics and 3j-symbols,
where $l,l'$ and $m,m'$ are total angular momentum and pertinent magnetic
quantum numbers, respectively~\cite{hwg97}:

\begin{eqnarray}
&& M^{jj'}_{lm,l'm'} =\delta^{jj'}\delta_{ll'}\delta_{mm'}
                 + (1-\delta^{jj'})\,
i^{2m+l'-l}\,\sqrt{4\pi(2l\!+\!1)(2l'\!+\!1)}\nonumber\\
&&\qquad\mbox{}\times
\left({\frac{a_j}{a_{j'}}} \right)^2
{\frac{{ j_l(k a_j)}}{{ h_{l'}^{(1)}(k a_{j'})}}}
\sum_{l'' =0}^{\infty}\sum_{m''=-l'}^{l'}
\sqrt{{2l''}+1}
 \,i^{l''}\,
{\left({\begin{array}{ccc}
  {l''}& {l'}&{l} \\
  {0}&{0}&{0} \end{array}}\right)}\nonumber\\
&&\qquad\mbox{}\times
 {\left({\begin{array}{crc}
  {l''} & {l'} & {l}\\
  {m\!-\!m''} & {m''} & {-m} \end{array}}\right)}
D^{l'}_{m'\!,m''}(j,j')\,
{  h_{l''}^{(1)}(k r_{jj'})}\,
{ Y_{l''}^{m\!-\!m''}\bigl( \hat r^{(j)}_{jj'}\bigr)}.
\label{Msphere}
\end{eqnarray}
The unit vectors $\hat r^{(j)}_{jj'}$  
point from the origin of sphere $j$ (as measured in its local
coordinate system) to the origin of sphere $j'$. The
local coordinate system of sphere $j'$ is mapped  to the one of sphere
$j$ with the help of the rotation matrix
$D^{l'}_{m'\!,\tilde m}(j,j')$.

For small scatterers the expression of the multi-scattering matrix even 
simplifies:
\begin{equation}
M^{jj'}(\veps)\approx
\delta ^{jj'} - \bigl(1-\delta ^{jj'}\bigr)
{{f_j^{s}(\veps)}}  
 \frac{ \exp (ik{r_{jj'}})}{{r_{jj'}}}
\, \ + \ \, {\cal O}(\mbox{$p$-wave})\,.
\end{equation}
since only $s$-wave scattering is important;
i.e., 
spherical waves modulated by  $s$-wave amplitudes $f_j^{s}(E)$ 
propagate between
the spheres. The integrated d.o.s.\ in the
case of two small spherical cavities of common radius $a$ and center-to-center
separation $r$ is~\cite{abaw01} 
\begin{equation}
{\cal{N}}_{\rm C}^{\rm oo}(\veps )=-\frac{1}{\pi}{\rm Im}\ln\det
M^{{\rm oo}}(\veps)\approx \nu_{\rm deg}
\frac{{ a^2}}{\pi { r^2}} \sin [ 2({ r}-{ a})k]
+ {\cal{O}}\left((ka)^3\right)
\end{equation}
where $\nu_{\rm deg}$ is the spin/isospin--degeneracy factor.
This expression should be compared with the semiclassical approximation
that sums up all partial waves
\begin{equation}
{\cal{N}}_{\rm C,sc}^{\rm oo}(\veps )= \nu_{\rm deg}
\frac{a^2}{4\pi r(r-2a)} \sin [ {{2(r-2a)}k}]\,.
 \label{Gutztrace}
\end{equation}
In fact, the latter  is  the leading contribution to Gutzwiller's trace
formula~\cite{gutbook}, namely the non-repeated
contribution of  the two-bounce periodic orbit
between the two spheres, with the action
$S_{po}(k)/\hbar = 2(r-2a)k$  where $2(r-2a)$ is the length of the
geometric path. Note that the
semiclassical result is suppressed by a
factor of $1/4$ in comparison to the small-scatterer one.

As shown in Ref.\,\cite{abaw01} the semiclassical result is a very good
approximation of the full quantum mechanical
result calculated from the exact expression
(\ref{Msphere}) of the two-sphere scattering matrix when plugged into
the modified Krein formula (\ref{modKrein}).
Therefore the Casimir energy for the two spherical cavities inside a
non-relativistic non-interacting fermion background can be approximated
in terms of a spherical Bessel function $j_1$ as
\begin{eqnarray}
\Ec^{\rm oo}\ =\ - \int_0^{\mu(k_F)} d E\, {{\cal N}_{\rm C}^{oo}(E)}
\approx -\nu_{\rm deg}\, \mu \frac{a^2}{2\pi r(r-2a)} j_1[2(r-2a)k_F]\,,
\end{eqnarray}
which is valid for $k_F a >1$.
This expression is long-ranged, i.e.\ $1/L^3$ with $L=r-2a$.
For the sphere-plate system the Casimir energy reads instead
\begin{eqnarray}
\Ec^{{\rm o} {\mathbf \vert}}
\approx -\nu_{\rm deg}\, \mu \frac{a}{2\pi (r-a)} j_1[2(r-a)k_F]\,,
\end{eqnarray}
which scales even as $1/L^2$ with $L= r-a$.
Note that in both cases, the two-sphere system or the sphere-plate system,
the Casimir
energy does not have a fixed sign in contrast
to the standard Casimir effect with
fluctuating electromagnetic or scalar fields between these obstacles. Instead
the sign of the Casimir energy oscillates as function of the action of the
two-bounce orbit. Therefore, under  increase of
the distance between the cavities,
the Casimir energy, which starts out to be attractive, can be made repulsive,
and under a further increase of this distance, it can become
attractive again,
with a decreased strength of course. The reason for this new type of behaviour
of a Casimir energy is the presence of a new scale in addition to the length
scales, namely the chemical potential $\mu$.
In fact,
the strength of this fermionic Casimir energy scales with the strength of
the chemical potential and therefore with the UV-cutoff of the theory. Also
this behavior distinguishes the fermionic Casimir effect from the standard
Casimir effect: the latter is governed by the infrared behavior of the
corresponding density of states.

\section{Fermionic Casimir effect between superfluid grains}
 
\begin{figure}[h,t,b]

\centerline{\includegraphics[height=3cm]{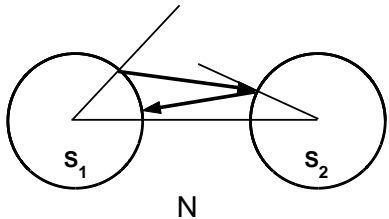}}

\caption{ \label{fig:andreev}
S$_1$ and S$_2$ are two superfluid grains immersed in a normal Fermi 
liquid~N.  
Indicated is  the Andreev reflection between an incoming particle (hole) 
and an outgoing hole (particle). The figure is from Ref.~\cite{bmw05}.}
\end{figure}

A generalization of the fermionic Casimir effect to the case of superfluid
cavities
immersed in normal fermionic matter is reported in 
Ref.\,\cite{bmw05}.
The pairing interaction due to the superfluid obstacles 
leads to enhanced 
Casimir contributions, because of the dominance of the particle-hole terms
over the particle-particle and hole-hole contributions. Semiclassically, this
can be explained by the focusing (and only for large separations
defocusing)
nature of the Andreev reflections~\cite{andreev} (see Fig.~\ref{fig:andreev})
in comparison to the defocusing specular reflections
at normal circular boundaries. 
Since the particle-hole d.o.s\
has the opposite sign to the particle-particle one and since  the d.o.s\
has
to be integrated over the quasi-particle energy as measured from the
Fermi surface, the resulting Casimir energy for superfluid obstacles 
is strongly repulsive~\cite{bmw05}.

\section{Application of the map method to the scalar Dirichlet problem} 
\begin{figure}[h,t,b]

\psfrag{log2(L/a)}{$\log_2\, (L/a)$}
\psfrag{-E/hbar c pi**3/1440 L**2}
{${\cal E}_{\rm C}(a,L)\, /\,(- \frac{\hbar c\pi^3 a}{1440 L^2})$}

\centerline{\includegraphics[width=8.8cm,angle=-90]{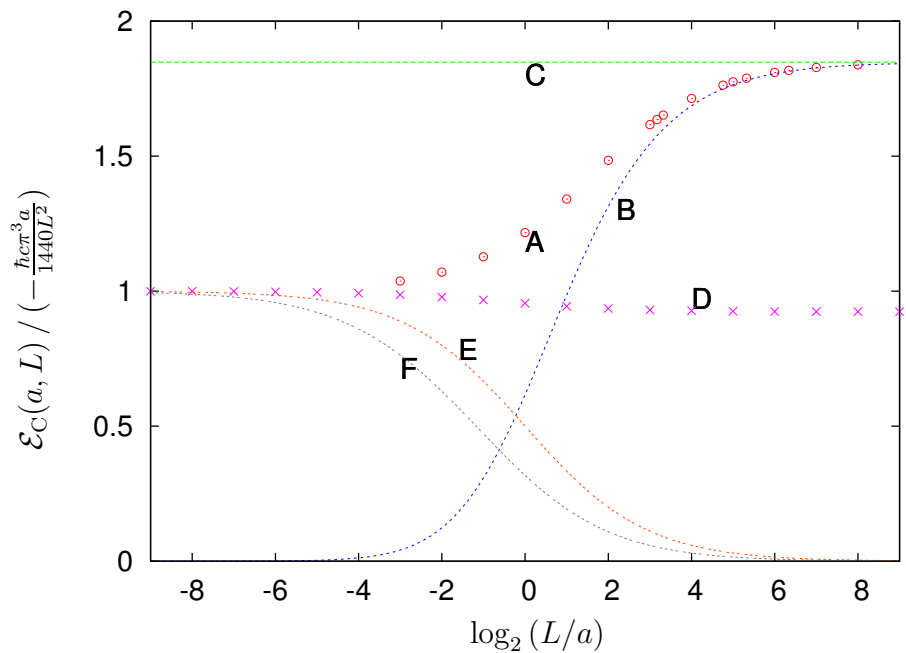}}

\caption{Predictions for the scalar Casimir energy
${\cal E}_{\rm C}(a,L)$ of the Dirichlet sphere-plate configuration
in units of $-1400\hbar c \pi^3 a/L^2$
as function
of the distance $L$ divided by the radius $a$ of the sphere.
The circles (A) represent the numerically calculated exact expression 
for the sphere-plate system, the curve (B) shows the $s$-wave approximation, 
the line (C) represents the asymptotic limit $2\times 90/\pi^4$,
the crosses (D) represent the numerically calculated results from 
the semiclassical Gutzwiller  formula including {\em all} repeats, 
see \cite{bmwscalar}.
The curves (E) denotes the plate-based, the curve (F) 
the sphere-based proximity-force approximation,
see \cite{gies,scar}.
 }
\label{fig:E_C-plots}
\end{figure}

Note that the map method onto a scattering problem
can be generalized from the Casimir effect
in the Fermi sea to other systems; it is especially applicable for
the case  of a fluctuating scalar field between
two spheres or a sphere and a plate (with Dirichlet boundary
conditions)~\cite{gies,scar}.
Whereas the $s$-wave approximation in the small-scatterer limit is not 
important for the fermionic Casimir energy which 
is governed by the UV part of
the d.o.s.\ (i.e.\ by the contribution at  the chemical potential),
it is essential for the
large distance physics of the fluctuating-scalar
Casimir effect, as the latter is governed by the infrared behaviour
of the d.o.s. Remember the relative factor of four between the $s$-wave
and semiclassical result of the two-cavity d.o.s. The same expressions
are valid for the d.o.s.\ of the fluctuating-scalar Casimir effect. Therefore,
the Casimir energy for two far separated
Dirichlet spheres is enhanced by
a factor of $4\times (90/\pi^4)$  \cite{bmwscalar} relative 
to the result of the proximity force approximation
which is only valid for small separations~\cite{proximity}. 
The extra $90/\pi^4$ takes into 
account the missing repeats of the two-bounce orbit in the 
semiclassical formula
(\ref{Gutztrace}), which are unimportant at large separations, but needed 
at small ones (see curve D of Fig.~\ref{fig:E_C-plots}).
For the sphere-plate configuration
the corresponding enhancement at large separations
is  $2\times 90/\pi^4$, because  only one
sphere is involved; see Fig.~\ref{fig:E_C-plots} for the comparison of this
result (C) with  the
exact calculation (A)
and the $s$-wave approximation (B)~\cite{bmwscalar}  
and with 
the proximity force approximations  of Refs.~\cite{gies,scar}.  
The exact data (A)
are compatible with numerical data of the wordline
approach, within error bars, \cite{gies} 
in the window of overlap, i.e.\ $L\leq 4a$.
In the case of the electromagnetic Casimir effect, the
$s$-wave dominance at large separation has to be replaced by a
$p$-wave dominance, since the charge-neutrality of the sphere forbids a
monopole
term, whereas the standard Casimir-Polder energy is  dominated by 
induced-dipole contributions.

\section{Summary}
We have shown that  there exists an effective interaction between voids
inside a Fermi gas background, even if the fermions are non-interacting.
This new form of Casimir energy is neither attractive nor repulsive,
but oscillates  according to the relative arrangement of  the cavities
(in the superfluid case it is strongly repulsive).
The spheres can be replaced with other objects, if the curvature radii are
larger than the Fermi wave length.
The effects of finite surface thickness can be booked as Weyl-term
contributions
and do not affect the geometrical Casimir part
if the objects do not overlap.
Since the Casimir interaction between bubbles
is oscillating and rather long-ranged (in comparison to van der Waals terms),
disordered lattices are expected as emerging structures.
The disorder can be further enhanced by finite temperature $T$ and corrugated
surfaces. 
The map method can be generalized to other Casimir systems,
especially to the scalar Casimir effect for Dirichlet spheres or sphere-plate
systems.

\ack
A.W. would like to thank Professors Emilio Elizalde, Sergei Odintsov and
Joan Soto
for the excellent organization of the Workshop QFEXT'05 and 
Holger Gies and Antonello Scardicchio for discussions.
Support from
the Forschungszentrum J\"ulich under contract No.
41445400 (COSY-067), 
from 
the Department of Energy under grant DE-FG03-97ER41014, and from the
Polish Committee for Scientific Research (KBN) under Contract
No.~1~P03B~059~27  
is gratefully acknowledged.

\end{document}